\documentstyle[aps,prl,twocolumn,epsf,floats]{revtex}

\newcommand{\equ}[1]{(\protect\ref{#1})} 
\newcommand{\Journal}[4]{#1 {\bf #2}, #3 (#4)}

\newcommand{\al}{\alpha} \newcommand{\la}{\lambda}
\newcommand{\ep}{\varepsilon} \newcommand{\Par}{\parallel}
\newcommand{\Per}{\perp} \newcommand{\ptime}{\partial_t}
\newcommand{\ppar}{\nabla_\parallel} 
 \newcommand{\xpar}{x_\parallel}
 \newcommand{\eperp}{{\bf e}_\perp}
\newcommand{\epar}{{\bf e}_\parallel}

\begin{document}

\draft

\wideabs{

\title{Stochastic equation for the erosion of inclined topography} 

\author{Romualdo Pastor-Satorras and Daniel H. Rothman}

\address{Department of Earth, Atmospheric, and Planetary Sciences\\
Massachusetts Institute of Technology, Cambridge, Massachusetts 02139}

\maketitle 

\begin{abstract} 
We present a stochastic equation to model the erosion of  topography
with fixed inclination.  The inclination causes the erosion to be
anisotropic.  A zero-order consequence of the anisotropy is the
dependence of the prefactor of the surface height-height correlations
on direction.  The lowest higher-order contribution from the
anisotropy is studied by applying the dynamic renormalization group.
In this case, assuming an inhomogenous distribution of soil material,
we find a one-loop estimate of the roughness exponents.  The predicted
exponents are in good agreement with  new  measurements made from
seafloor topography.
\end{abstract}

\pacs{PACS numbers: 92.40.Gc, 64.60.Ht, 05.60.+W, 64.60.Ak}

}

The rich complexity of the Earth's surface,  both on land and beneath
the sea, is the result of physical mechanisms ranging from tectonic
motion to surficial erosion \cite{scheidegger91,iturbe97}.  Despite
this variation, however, geologic surfaces show a certain degree of
universality: they may often be characterized as {\em self-affine}
\cite{mandelbrot82,surfaces} over some range of length scales. This
means that, if $h(\vec{x},t_0)$ is the  height of the surface at
position $\vec{x}$ at some time $t_0$, then the ``roughness'',
measured by 
the height-height static correlation function $C(\vec{x}) =
\langle  ( h(\vec{x}, t_0) - h(0, t_0) )^2 \rangle^{1/2}$, grows as
$x^\alpha$, where $\alpha$ is called the {\em roughness exponent}
\cite{surfaces}.  Empirical measurements of $\alpha$ are numerous.
While many indicate that $\alpha$ is small ($0.30 < \alpha < 0.55$)
\cite{small_alpha_refs,crossover}, a number of other  measurements
show it to be large ($0.70 < \alpha < 0.85$)
\cite{crossover,large_alpha_refs,czirok93}.  Moreover, some
measurements indicate that $\alpha$ crosses over from large to small
values as length scales become greater than approximately 1 km
\cite{crossover}.  Motivated by these findings, we propose that the
large values of $\alpha$ at small length scales may be explained by the
influence of a preferred direction---downhill---for  the flux of
eroded material.  We derive an anisotropic noisy diffusion equation to
describe erosion at the small length scales where the preferred
direction is fixed throughout space.  Under the additional assumptions
that the flux of eroded material increases with increasing distance
downslope and that the dominant effects of noise are fixed in space,
we find,  using  the dynamic renormalization group (DRG), a first-order
estimate of the roughness exponents.  New measurements of our own,
made from  the topography of the continental slope off the coast of
Oregon, are  in good agreement with our predictions. 
We find that our anisotropic theory significantly enriches previous isotropic
continuum models \cite{sornette93,many} for two reasons.
First, it predicts that correlations differ in different directions, and
and second, it predicts that these correlations decay quantitatively 
differently than they do for isotropic topography.

\begin{figure}
\epsfxsize=8truecm    \centerline{\epsfbox{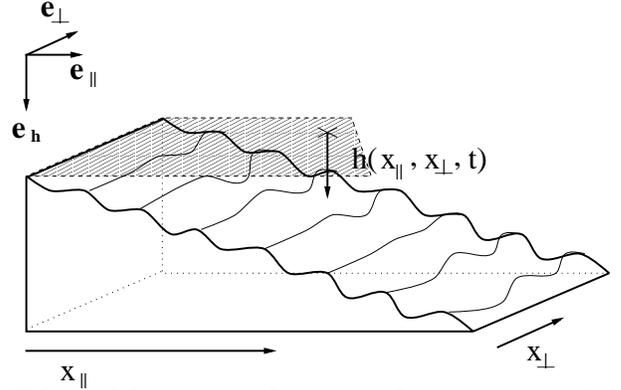}}
\caption{Schematic configuration of an anisotropic landscape for the
case $d=2$.}
\label{landscape}
\end{figure}

Figure \ref{landscape} depicts the framework for our theory: a surface
$h$ on a two-dimensional substrate.  We refer to $h$ generically by
the term {\em landscape}, and note that it inclination is fixed.  The
unit vector ${\bf e}_h$ is the ``growth'' direction, which  is
measured downwards from the top of the slope.  The preferred,
downhill, direction is given by the unit vector $\epar$, while
$\eperp$ represents a vector perpendicular to $\epar$ and ${\bf e}_h$.
Later, when applying the DRG, we will generalize to  landscapes on a
$d$-dimensional substrate; in this case $\eperp$ represents the
subspace of all directions  perpendicular to $\epar$ and ${\bf e}_h$,
and has dimension $d-1$.  The configuration is completed by selecting
fixed boundary conditions at the top of the slope, $\xpar=0$, or by
imposing the symmetry $\xpar\to-\xpar$.

Due to the preferred direction $\epar$ in Fig.~\ref{landscape},   the
statistical properties of $h$ may be anisotropic.  Thus, if $h$ is
self-affine, we expect different roughness exponents for
correlations measured in each of the directions $\epar$ and $\eperp$.
Thus we define $\al_\Par$ and $\al_\Per$ such that $C_\Par(x_\Par)
\sim x_\Par^{\al_\Par}$ for correlations  along a fixed transect
$\vec{x}_\Per^0$ = const., and  $C_\Per(\vec{x}_\Per)  \sim
x_\Per^{\al_\Per}$ for correlations  along a fixed transect  $x_\Par^0
=$ const., where  in general $\al_\Par \neq \al_\Per$.  These
relations can be summarized in the single scaling form 
\begin{equation}
C(x_\Par, \vec{x}_\Per)  \sim  b^{\al_\Par}  C(b^{-1} x_\Par,
b^{-\zeta_\Par} \vec{x}_\Per),
\label{scalaniso}
\end{equation}
where $\zeta_\Par$ is the {\em anisotropy exponent}. The exponents
$\al_\Par$  and $\al_\Per$ are related through $\al_\Per = \al_\Par /
\zeta_\Par$.  The exponent $\zeta_\Par$ accounts for the different
rescaling factors along the two main directions. Since the space is
anisotropic, when performing a scale change, we must rescale $x_\Par$
and $\vec{x}_\Per$  by different factors $b_\Par$ and $b_\Per$,
respectively, if we are to recover a surface with the same statistical
properties.  We assume in our model that $\zeta_\Par = \log b_\Per /
\log b_\Par$ = const. 

We seek a single stochastic equation for the landscape height $h$.
Whereas others \cite{sornette93} have advocated the now classical,
isotropic, non-conservative interface growth 
equation due to Kardar, Parisi, and Zhang (KPZ) \cite{kardar86},
we assume  here that the underlying soil is locally conserved
such that
\begin{equation}
\ptime h = - \nabla \cdot \vec{J} + \eta,
\label{firstaniso}
\end{equation}
where $\vec{J}$ is the current of soil per unit length. The soil
however is not globally conserved, since it is lost at the bottom
boundary.  We also allow local conservation to be broken by the
addition of a stochastic noise term $\eta$, discussed below.

Physically, the current $\vec{J}$ is expected to reflect two
effects. First, we expect a  local isotropic diffusing component,
tending to smooth out the surface.  Second, we expect an average
global flow  of dragged soil, directed mainly downhill.   Thus we
postulate the following form  for the current:
\begin{equation}
\vec{J} = -\nu \nabla h - \Gamma \ppar h.
\label{soilcurrent}
\end{equation}
The first term corresponds to Fick's law for diffusion, and represents
the isotropic relaxational dynamics of the soil.  The second term
represents the average flow of soil that is dragged downhill, either
due to the flow of water or the scouring of the surface by the flow of
the soil itself.  The direction of this term is given by the vector $
\ppar h \equiv \partial_\Par h \: \epar $.  The term $\Gamma$ plays
the role of an {\em anomalous anisotropic diffusivity}.  In order to
gain insight into the role of $\Gamma$, consider the case in which
erosion results from the stress exerted on the soil bed by an overland
flow $q$ of water, where $q$ is the volumetric flow rate though unit
area perpendicular to the direction of steepest descent.  The greater
$q$ is, the stronger is the stress \cite{scheidegger91,iturbe97}.
Moreover, since $q$ flows downhill, it increases with distance
downslope. Thus $\Gamma$ must be an increasing function of $\xpar$.
Since the fixed inclination implies that $h$ increases with $\xpar$,
we choose to parameterize the anomalous diffusion as a function of the
height such that $\Gamma\equiv\Gamma(h)$ \cite{pelletier97}.  Defining
$\Gamma(h)=\la_0 + g(h)$, with $g(0)=0$ and $G(h)=\int g(h) dh$, we
substitute Eq.~\equ{soilcurrent} into \equ{firstaniso}. 
Since $g(h)\partial_\Par h = [dG(h)/dh ] \partial_\Par h
  = \partial_\Par G(h)$, 
where we have used the chain rule for the second equality, we obtain
\begin{equation}
\ptime h = \nu_\Par \partial_\Par^2 h + \nu_\Per \nabla_\Per^2 h +
\partial_\Par^2 G(h) + \eta,
\label{interaniso}
\end{equation}
where $\nu_\Per=\nu$ and $\nu_\Par=\nu+\la_0$.

Even in the absence on any nonlinearity, fundamental conclusions may
be drawn from \equ{interaniso}. By setting $g=0$ (i.e., by considering
$\Gamma(h)=\la_0\equiv$ const.), we obtain a linear equation which is
an anisotropic  counterpart of the Edwards-Wilkinson equation
\cite{surfaces}.
In can then be easily shown \cite{surfaces} that the correlation
functions along the main directions $\epar$ and $\eperp$ are inversely
proportional to the square root of the diffusivities $\nu_\Par$ and
$\nu_\Per$ respectively, that is, $C_\Per / C_\Par \sim
(\nu_\Par/\nu_\Per)^{1/2}$.  In other words, since the preferred
direction gives $\nu_\Par > \nu_\Per$, the topography is
quantitatively rougher, at all scales and by the same factor, in the
perpendicular direction than in the parallel direction.

In order to obtain more information on the scaling properties of
Eq.~\equ{interaniso}, we have studied it using the DRG.  Assuming that
$\Gamma(h)$ is an analytical function,  we can perform a Taylor
expansion in powers of $h$.  Since all odd powers of $h$ must vanish
in order to the preserve the joint  symmetry $h\to-h$,
$\vec{J}\to-\vec{J}$ in Eq.~\equ{firstaniso}, we are left at lowest
order with $g(h) \simeq \la_2 h^2$. By dimensional analysis  one can
check that all the terms in this expansion  are relevant 
under rescaling.  However, the flux $Q(\xpar)$ of the erosive agent
(water or 
soil)  flowing on the surface should grow no faster than  $Q(\xpar)
\sim \xpar^{d}$.  Then, taking $h \sim \xpar$, we find that the terms
in $g(h)$ should be of order $h^{d}$ or less.  Specializing to the
case of $d=2$ (i.e., real surfaces), we then find it reasonable to
truncate $g$ at second order, such that  Eq.~\equ{interaniso} takes
the form
\begin{equation}
\ptime h = \nu_\Par \partial_\Par^2 h + \nu_\Per \nabla_\Per^2 h +
\frac{\lambda}{3}  \partial_\Par^2 (h^3) + \eta,
\label{lastaniso}
\end{equation}
where $\la=\lambda_2$.  
Note that Eq.~\equ{lastaniso} differs from
the anisotropic driven diffusion equation of Hwa and Kardar
\cite{hwa92} because the form of our current $\vec{J}$ is suggested
not only by symmetry arguments, but also by the physics of erosion.

We now address the issue of noise.  We distinguish two different
sources.  First, we may allow a term of ``annealed'' noise,
$\eta_t(\vec{x}, t)$,
depending on time and position, and  
describing a random, external forcing, due to, for example,
inhomogeneous rainfall.
We  assume that this noise is
isotropic, Gaussian distributed, with zero mean, and  uncorrelated
such that $\left< \eta_t(\vec{x},t) \eta_t(\vec{x}',t') \right> = 2 D_t
\delta^{(d)}(\vec{x}-\vec{x}') \delta(t-t')$.  Second, we may have a
term of 
``quenched'' noise to account for the heterogeneity of the soil,
mimicking the variations in the erodibility of the landscape
\cite{czirok93}.  We  represent this randomness by a source of
Gaussian {\em static} noise $\eta_s(\vec{x})$, with correlations $\left<
\eta_s(\vec{x}) \eta_s(\vec{x}') \right> = 2 D_s
\delta^{(d)}(\vec{x}-\vec{x}')$. This form of 
noise has  been previously proposed to model soil heterogeneity in
cellular automata models of fluvial networks \cite{quenched}.  In the
following we consider  the limits (i) $\eta_s=0$ ($D_s \ll D_t$),
corresponding to a situation of random external forcing  and
homogeneous 
composition of soil, and (ii) $\eta_t=0$  ($D_s \gg D_t$),
representing the limit in which the external forcing is constant and
the most essential source of noise is the inhomogeneous composition of
the soil. 

Application of the DRG follows the procedure used in
Refs.~\cite{hwa92,corral97}.  In Fourier space we proceed by
integrating over the shell of large wave vectors $\Lambda e^{-l} < k <
\Lambda$, where $\Lambda$ is the wave vector upper cutoff and $e^l$ is
the rescaling factor, and by subsequently rescaling the system back to
its original size through the transformation $\vec{x}_\Per \to e^l
\vec{x}_\Per$,  $\xpar \to e^{l\zeta_\Per} \xpar$, $h \to
e^{l\al_\Per} h$, and $t \to e^{l z_\Per} t$. The anisotropy is
explicitly included in the exponent $\zeta_\Per \equiv
\zeta_\Par^{-1}$.   To lowest order in perturbation theory, both
limits (i) and (ii) above provide the same form for the
renormalization  group flow equations: 
\begin{eqnarray}
&& \frac{d \nu_\Par}{d l} = \nu_\Par (z_\Per - 2 \zeta_\Per +
\bar{\la}_i) ,\qquad \frac{d \nu_\Per}{d l} = \nu_\Per (z_\Per - 2)
\nonumber \\  && \frac{d \la}{d l} = \la (z_\Per + 2 \al_\Per - 2
\zeta_\Per - \frac{3}{2} \bar{\la}_i) \nonumber \\   && \frac{d D_i}{d
l} = D_i (\kappa_i z_\Per - 2 \al_\Per - \zeta_\Per - d + 1),
\nonumber 
\end{eqnarray}
where $i=t, s$ stands for the limits (i) and (ii) above,
respectively. Here $\bar{\la}_i$ is an effective coupling constant,
depending on the type of noise:  $\bar{\la}_t=\la D_t
K_{d-1}\Lambda^{d-2} / 2 \nu_\Par^{3/2} \nu_\Per^{1/2}$  in  (i), and
$\bar{\la}_s=\la D_s K_{d-1}\Lambda^{d-4} / 2 \nu_\Par^{3/2}
\nu_\Per^{3/2}$ in  (ii),  with  $K_d = S_d /(2\pi)^d$ and $S_d$ the
surface area of a $d$-dimensional unit sphere. The value of the
correction factor $\kappa_i$ is $\kappa_t=1$ and $\kappa_s=2$.  The
flow equations for $\nu_\Per$ and $D_i$ are exact to all orders in the
perturbation expansion \cite{hwa92,lai91}. 
They provide us with the exact result $z_\Per=2$ \cite{note}.
The effective coupling flows under rescaling
as
\begin{equation}
\frac{d \bar{\la}_i}{d l} =\bar{\la} _i (\ep_i - 3 \bar{\la}_i),
\label{floweq}
\end{equation}
where $\ep_i = d_c^{(i)} - d$, and $d_c^{(i)}$ is the critical
dimension for each particular limit, $d_c^{(t)} = 2$ and $d_c^{(s)} =
4$. The stable fixed points of \equ{floweq} are $\bar{\la}^*_i = 0$
for $d>d_c^{(i)}$ and $\bar{\la}^*_i = \ep_i / 3$ for
$d<d_c^{(i)}$. For $d>d_c^{(i)}$ the critical exponents attain in both
limits their mean-field values   $\al_\Per^{MF}=0$, $\zeta_\Per^{MF} =
1$, and $z_\Per^{MF} = 2$.  On the other hand, for $d<d_c^{(i)}$, the
critical exponents computed at first order in the $\ep$-expansion are:
\begin{equation}
\al_{\Per(i)}=\frac{5 \ep_i}{12}, \qquad \zeta_{\Per(i)} = 1 +
\frac{\ep_i}{6}.
\label{exponents}
\end{equation}

The physically relevant dimension for erosion is $d=2$. In the limit
of thermal noise this corresponds to the critical dimension. By
continuity, the exponents are $\al_\Per=\al_\Par=0$ and
$\zeta_\Per=\zeta_\Par = 1$. This result is consistent with a flat
landscape, with logarithmic corrections to the roughness
\cite{surfaces}.  However, we still expect anisotropy to appear in the
prefactor of the correlation functions $C_\Par$ and $C_\Per$, as
argued above.  On the other hand, in the limit of static noise we are
below the critical dimension, and \equ{exponents} is applicable.
Substituting $\ep_s=2$ we obtain the roughness exponents
\begin{equation}
\al_\Per = \frac{5}{6} \simeq 0.83, \qquad \al_\Par=
\frac{\al_\Per}{\zeta_\Per} = \frac{5}{8} \simeq 0.63. 
\label{numpredict}
\end{equation}

The values \equ{numpredict} predicted for $\al_\Per$ and $\al_\Par$
are in reasonable agreement with previous measures made at small
length scales \cite{crossover,large_alpha_refs}.  However, these
measurements were either averaged over all directions or the direction
of the measurements was not reported.  Thus, to check our results with
a natural landscape that has an unambiguous preferred direction, we
have analyzed digital bathymetric maps of the continental slope off
the coast of Oregon.  In this case the slope results from the
relatively abrupt increase in the depth of the seafloor as the
continental shelf gives way to the deeper continental rise.  Figure 2a
shows one portion of this region.  Here the main feature of the
topography is a deep incision called a {\em submarine canyon}.  In
this region, submarine canyons are thought to have resulted from
seepage-induced slope failure \cite{orange94}, which occurs when
excess pore pressure within the material overcomes the gravitational
and friction forces on the surface of the material, causing the slope
to become unstable.  Slope instabilities then create submarine
avalanches, which themselves can erode the slope as they slide
downwards.

\begin{figure}[t]
\epsfxsize=8truecm    \centerline{\epsfbox{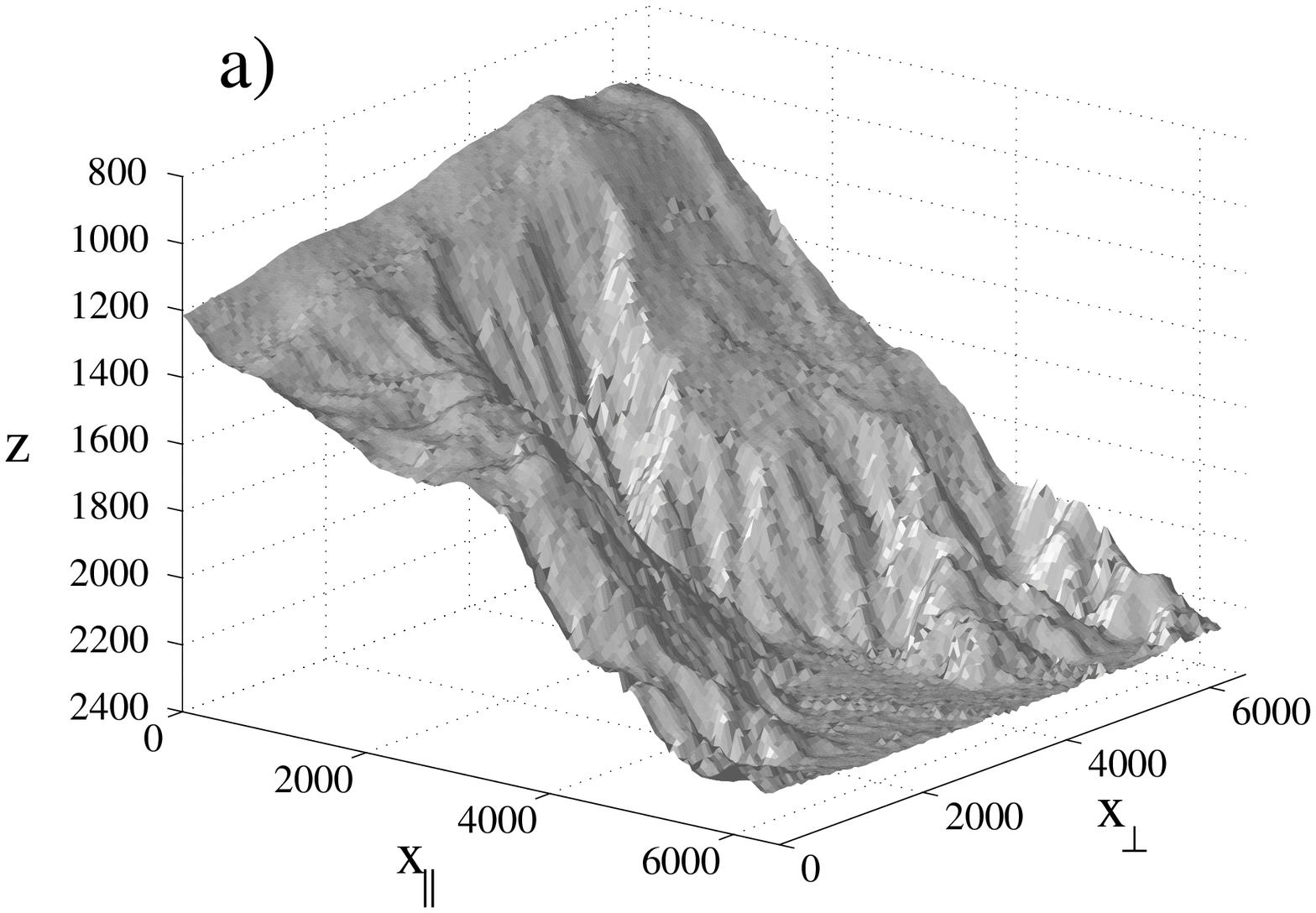}}
\epsfxsize=8truecm    \centerline{\epsfbox{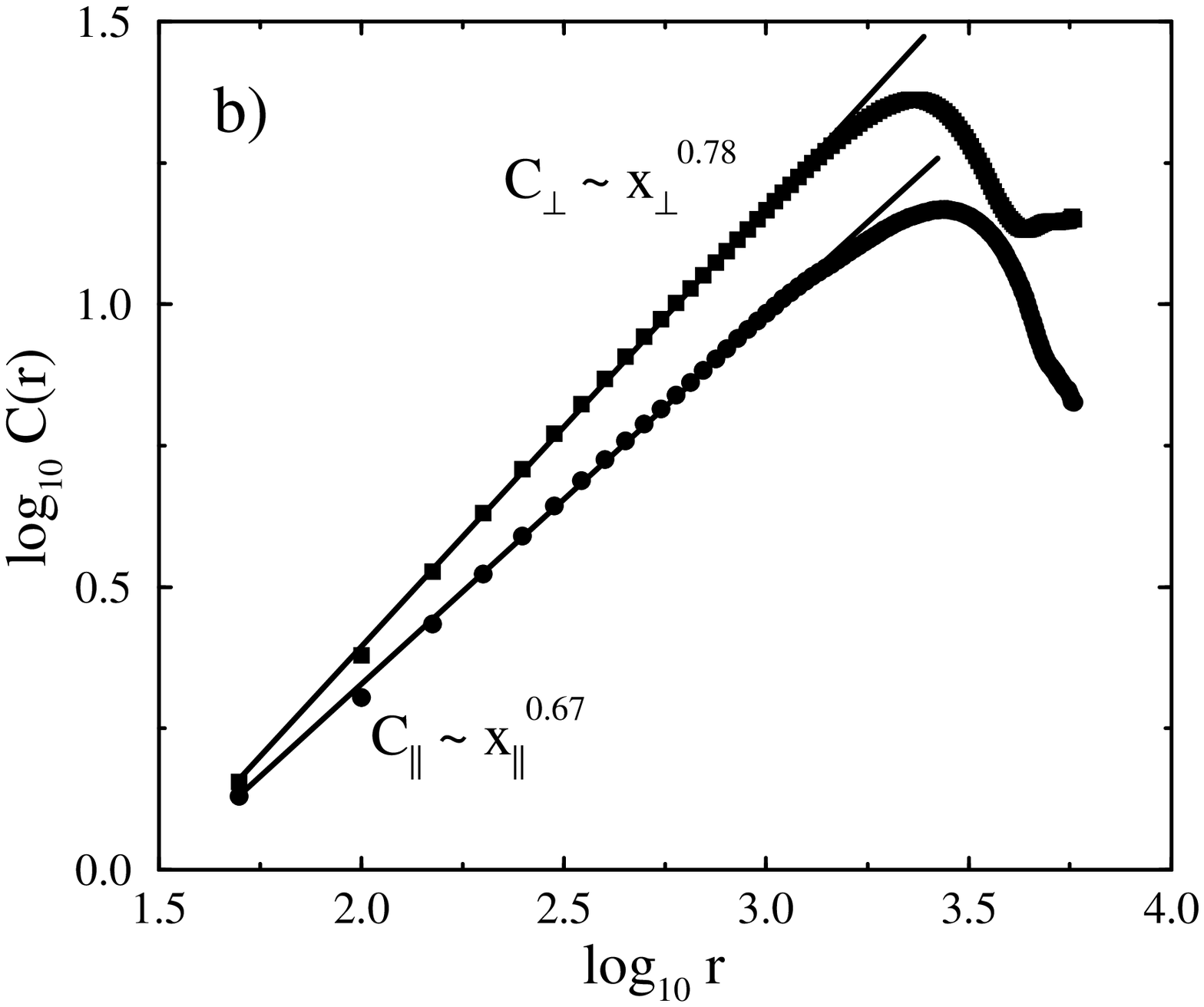}}
\caption{
(a) Digital map of a submarine canyon off the coast of Oregon,
located at coordinates $44^o 40'$ N,
$125^o 45'$ W. The vertical axis represents depth below
sea level. Distances are measured in meters. 
(b) Height-height correlation functions computed 
along the parallel ($C_\Par$) and perpendicular ($C_\Per$) 
directions. Solid lines are least-squares fits to the scaling
region.}  
\label{correlations}
\end{figure}

Figure 2b shows the height-height correlation functions $C_\Par$ and
$C_\Per$, corresponding, respectively, to the parallel and
perpendicular directions of the seafloor topography in Fig.~2a.  
The computation of $C_\Per$ follows from its definition but
the computation of $C_\Par$ requires some comment.
The fluctuations measured by $C_\Par$ must be defined with respect to an
appropriate average profile. 
Briefly, one expects that geologic processes
other than erosion (e.g., tectonic stresses) are responsible for 
long-wavelength deformation in the parallel direction. 
We may estimate such systematic corrections by computing the mean 
profile in the parallel direction:
$h_{\rm av} (x_\Par) = L_{\perp}^{-1} \int dx_{\perp} h(x_\Par,x_\Per)$.
We then compute $C_\Par$ from the fluctuations of the detrended surface
$\tilde{h} = h - h_{\rm av} (x_\Par)$.
>From both $C_\Par$ and $C_\Per$ we find that
the least-squares estimates
of the roughness exponents, $\al_\Par \simeq 0.67$ and $\al_\Per
\simeq 0.78$, exhibit a surprisingly good fit to our theoretical
predictions \equ{numpredict}.

We have also measured $C_\Par$ and $C_\Per$ in some desert
environments.  In these cases (not shown), we did not obtain
conclusive power law scaling, but we always found $C_\Per / C_\Par >
1$, as predicted by the linear theory.  Thus, while the example of
Figure 2 may be in some sense specialized, one of our main
predictions---that the topography in the perpendicular direction is
rougher than the topography in the parallel direction---seems to be of
fairly general validity.

In conclusion, we note that the main elements of our theory are the
conservation of the eroded material, randomness of either the
landscape or the forcing, and the presence of a preferred direction
for the material transport.  The latter assumption leads to an
anisotropic equation that applies, in principle, to any erosive
process with the appropriate lack of symmetry. In the usual geological
setting, however, the anisotropy applies specifically to a surface of
fixed inclination which, in turn, implies that our theory should only
apply locally, to the relatively small scales where the preferred
direction of transport is approximately constant. Because the
anisotropy should vanish at large length scales, these large scale
features should be presumably described by an isotropic theory, such
as the KPZ equation \cite{sornette93,kardar86}. Indeed, the KPZ
equation predicts exponents that are approximately consistent with
large scale observations. Since these predictions differ from ours, it
may be possible to use our results to distinguish statistically
between features of the landscape due to erosion and features due to
larger-scale processes, such as tectonic deformation.

\vspace{0.5cm}

R.P.S. acknowledges financial support from the Ministerio de
Educaci\'on y Cultura (Spain).  The work of D.H.R. was supported in
part by NSF grant EAR--9706220.


\begin{references}
\bibitem{note}
The dynamic critical exponent $z$ measures the saturation length of
correlations as a function of time. However, since the time scales for
geomorphologic evolution are many orders of magnitude larger than
those possibly available for observation, speculations about the
actual value of $z$ are meaningless.
\end{references}
\end{document}